\documentclass[%
 reprint,
superscriptaddress,
amsmath,amssymb,
prb,
aps,
]{revtex4-2}

\usepackage{booktabs}
\usepackage{graphicx} 
\usepackage{dcolumn} 
\usepackage{bm} 
\usepackage{hyperref} 
\usepackage[]{siunitx} 
\usepackage{textcomp} 
\usepackage[english, capitalise]{cleveref} 
\usepackage{setspace} 
\usepackage{xcolor} 
\usepackage[english]{babel}
\definecolor{noteorange}{RGB}{255, 128, 0}
\definecolor{hellerot}{RGB}{255, 128, 128}
\definecolor{rosa}{RGB}{255, 192, 203}
\definecolor{green}{rgb}{0.0, 0.5, 0.0}
\usepackage{romannum}

\DeclareSIUnit{\flu}{\milli\joule\per\centi\meter\squared}
\DeclareSIUnit{\ps}{\pico\second}
\DeclareSIUnit{\fs}{\femto\second}
\usepackage[T1]{fontenc}

\begin{document}

\preprint{APS/123-QED}

\title{
Orientation-resolved ultrafast spin reorientation dynamics in ferrimagnetic DyCo$_5$
}
\author{Johanna Richter}
\email{johanna.richt3r@gmail.com}

\author{Martin Hennecke}
\email{hennecke@mbi-berlin.de}
\affiliation{%
 Max-Born-Institut für Nichtlineare Optik und Kurzzeitspektroskopie, Max-Born-Straße 2A, 12489 Berlin, Germany
}%
\author{Martin Schmidbauer}
\affiliation{%
 Leibniz-Institut für Kristallzüchtung, Max-Born-Straße 2, 12489 Berlin, Germany
}%
\author{Ilie Radu}
\affiliation{European XFEL, Holzkoppel 4, 22869 Schenefeld, Germany}
\author{Clemens von Korff Schmising}
\affiliation{%
 Max-Born-Institut für Nichtlineare Optik und Kurzzeitspektroskopie, Max-Born-Straße 2A, 12489 Berlin, Germany
}%
\author{Stefan Eisebitt}
\affiliation{%
 Max-Born-Institut für Nichtlineare Optik und Kurzzeitspektroskopie, Max-Born-Straße 2A, 12489 Berlin, Germany
}%
\affiliation{Technische Universität Berlin, Institut für Optik und Atomare Physik, 10623 Berlin, Germany}

\date{\today}

\begin{abstract}
Under quasi-static conditions, ferrimagnetic DyCo$_5$ thin films exhibit a thermally induced spin-reorientation transition, in which the equilibrium magnetization changes from an out-of-plane to an in-plane orientation, mostly as a result of the competing magnetic anisotropy contributions of the Dy $4f$ rare-earth and Co $3d$ transition-metal sublattices. 
While this equilibrium transition has been studied, it remains an open question whether such a reorientation can be triggered on ultrafast timescales using femtosecond laser excitation.
In this work, we investigate the ultrafast spin-reorientation dynamics in DyCo$_5$. 
The time-dependent orientation of the magnetization vector following femtosecond laser excitation is quantified by combining polar with transverse magneto-optical Kerr effect (MOKE) measurements in the extreme ultraviolet spectral range, which are sensitive to the out-of-plane and in-plane magnetization component, respectively. 
Both techniques are implemented at the Co M$_{3,2}$ resonance and are complemented by visible-light MOKE measurements. 
This combined approach allows us to resolve the canting of the magnetization from an out-of-plane toward an in-plane orientation and to determine the characteristic timescales of the transient spin-reorientation process.

\end{abstract}

\maketitle
\pagenumbering{arabic}
\section{Introduction}

All-optical magnetization switching in rare-earth and transition-metal (RE-TM) alloys has sparked interest in the ultrafast magnetization dynamics of multi-sublattice magnets~\cite{Stanciu2007,Radu2011,Kirilyuk_2013,hadri_materials_2017,scheid_light_2022}.
In particular, understanding magnetization switching processes, such as all-optical switching and the reorientation processes between out-of-plane and in-plane anisotropy states, is essential for future spintronic applications, such as optically controlled magnetic tunnel junctions or field sensors~\cite{wisniowski_magnetic_2012,chen_all-optical_2017}.

Among RE-TM ferrimagnets, DyCo$_5$ has emerged as a prototypical system for studying spin reorientation transitions (SRT). 
These transitions arise from competing anisotropy terms between sublattices with distinct temperature dependent magnetization and their exchange interactions~\cite{belov_spin-reorientation_1976,kimel_laser-induced_2004}. 
Early single-crystal magnetometry studies by \textcite{tsushima_spin_1983} identified two characteristic reorientation temperatures, $T_\mathrm{SR1}$ and $T_\mathrm{SR2}$, which delimit a rotational SRT region in which the equilibrium magnetization direction continuously changes with temperature \cite{ohkoshi_spin_1977}.

More recent work by \textcite{donges_magnetization_2017} employed a multiscale approach combining \emph{ab initio}-parametrized atomistic spin models with stochastic spin dynamics simulations in combination with x-ray magnetic circular dichroism (XMCD) measurements to reproduce the thermal magnetic behavior of DyCo$_5$. 
Their calculations capture the temperature dependence of the magnetization and demonstrate that the SRT originates from competing Dy and Co anisotropies with distinct thermal demagnetization characteristics. 
DyCo$_5$ also exhibits a ferrimagnetic compensation point at $T=\SI{148}{K}$, for which the net magnetization becomes zero. 
This phenomenon was investigated for varying compositions of DyCo alloys~\cite{iskhakov_effects_2004}.

While early element-specific XMCD measurements in the soft x-ray spectral range on DyCo alloys have provided first insights into their ultrafast demagnetization dynamics~\cite{abrudan_element-specific_2021}, the temporal evolution of the spin-reorientation has been elusive so far. 

In this work, we study the spin-reorientation dynamics in DyCo$_5$ by employing time-resolved magneto-optical Kerr effect (MOKE) measurements in the extreme ultraviolet (XUV) spectral range, which allow us to follow the evolution of the out-of-plane and in-plane magnetization components.
To this end, we combine the recently introduced polar geometry (XUV-P-MOKE)~\cite{richter_spectroscopic_2025} with the established transverse geometry (XUV-T-MOKE)~\cite{la-o-vorakiat_ultrafast_2009}, which is widely used as an element-specific probe of ultrafast demagnetization dynamics in reflection for in-plane magnetized multicompound systems~\cite{jana_experimental_2022,Korff_2023}. 
Complemented by visible-light (VIS) measurements in polar (P-MOKE) and longitudinal geometry (L-MOKE), we identify the characteristic timescales of the SRT.

\section{Experimental Methods}

\begin{figure} 
    \centering
    \includegraphics[width=\linewidth]{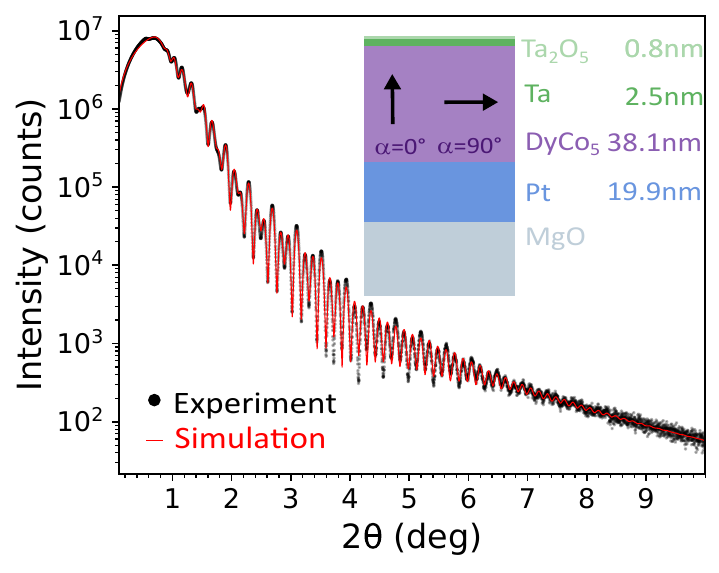}
    \caption{Hard x-ray reflectometry measurements compared to simulations for the determination of the sample layer thicknesses as indicated in the inset.}
    \label{fig:xraydiff}
\end{figure}

\begin{table}
    \centering
    \caption{Structural properties of the individual layers of the studied sample (thickness $t$, mass density $\rho$, rms roughness $r$) determined from the XRR measurements and simulations.}
    \label{tab:simulation_parameters}
    \begin{tabular}{lccc}
        \toprule
        Layer
        & t (\si{\nano\metre})
        & $\rho$ (\si{\kilo\gram\per\cubic\meter})
        & r (\si{\nano\metre}) \\
        \midrule
        MgO           & substrate      & $3580  \pm 500$ & $0.5 \pm 0.2$ \\
        Pt            & $19.9 \pm 0.5$ & $21021 \pm 500$ & $1.3 \pm 0.2$ \\
        DyCo$_5$      & $38.1 \pm 0.7$ & $9078  \pm 500$ & $0.9 \pm 0.2$ \\
        Ta            & $2.5 \pm 0.5$  & $16817 \pm 500$ & $0.5 \pm 0.2$ \\
        Ta$_2$O$_5$   & $0.8 \pm 0.5$  & $10143 \pm 500$ & $0.1 \pm 0.2$ \\
        \bottomrule
    \end{tabular}
\end{table}

Ferrimagnetic DyCo$_5$ thin films with the layer structure MgO($111$) / Pt(\SI{19.9}{nm}) / DyCo$_5$(\SI{38.1}{nm}) / Ta(\SI{2.5}{nm}) / Ta$_2$O$_5$(\SI{0.8}{nm}) were grown by magnetron sputtering. 
The layer thicknesses and densities were accurately determined via hard x-ray reflectometry (XRR) (see Fig.~\ref{fig:xraydiff}). 
The XRR measurements were performed on a \SI{9}{kW} Smartlab diffractometer system (Rigaku) using Cu K$\alpha_1$ radiation ($\lambda = \SI{1,54056}{\angstrom}$) with a primary collimation of better than $0.01^\circ$. 
Primary and secondary slits of \SI{0.1}{mm} have been selected to provide a resolution of the scattering angle $2\theta$ of about $0.02^\circ$. 
The corresponding simulations based on dynamical scattering theory were carried out using the software RCRefSimW \cite{Zaumseil_RCRefSimW}.
In the equilibrium state at room temperature, the DyCo$_5$ film studied  exhibits an out-of-plane magnetic anisotropy with a coercive field of approximately \SI{45}{mT}. 

\begin{figure} 
    \centering
    \includegraphics[width=\linewidth]{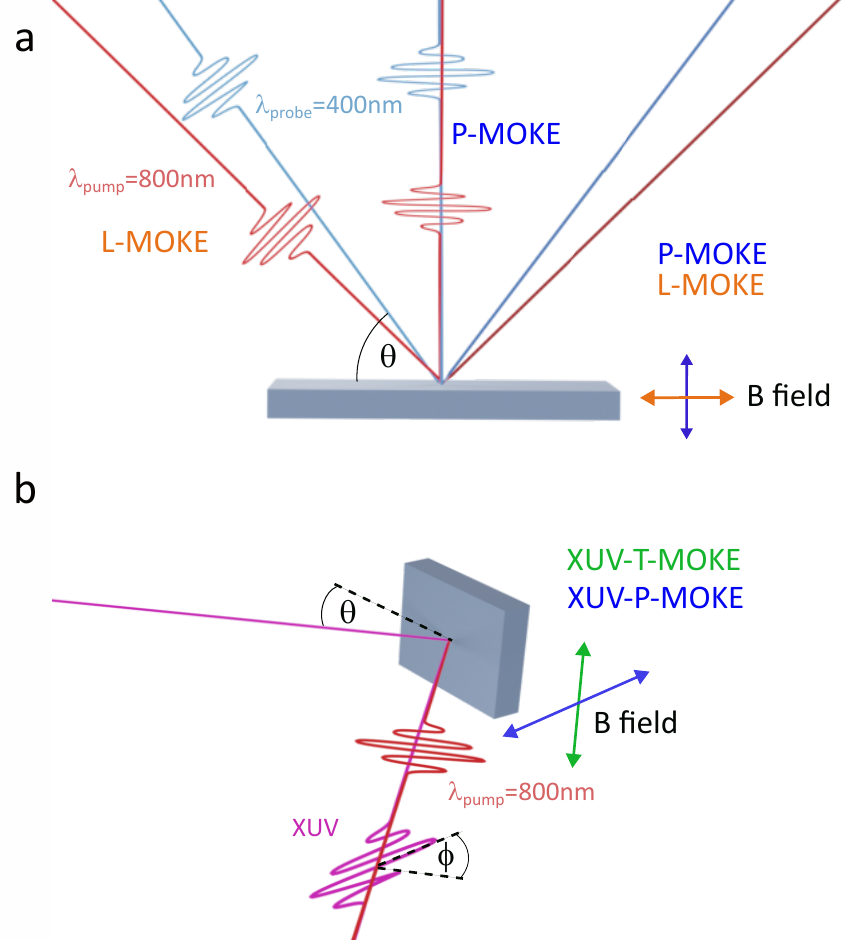}
    \caption{\textbf{a} VIS-MOKE geometries for polar (P-MOKE) and longitudinal MOKE (L-MOKE) \textbf{b} XUV-T-MOKE and XUV-P-MOKE geometries with incident probing polarization angle $\phi$. }
    \label{fig:setup}
\end{figure}

For the time-resolved VIS-MOKE measurements, optical excitation was provided by a Ti:sapphire laser system operating at a central wavelength of \SI{800}{nm}, while the probe pulses were frequency-doubled to a central wavelength of \SI{400}{nm}.
The overall temporal resolution was \SI{\approx 50}{fs}. 
Probing at a repetition rate of \SI{3}{kHz} with synchronized excitation at \SI{1.5}{kHz} allows consecutive detection of the pumped and unpumped state of the sample.
The measurements were carried out in polar and longitudinal geometry.
For P-MOKE, the sample was probed under normal incidence, while for L-MOKE, the probing angle was set to $\theta\approx\;$\SI{64.5}{\degree}.
In both cases, the pump pulses were incident on the sample almost collinearly with the probe pulses.
An alternating external magnetic field was applied either in-plane for the L-MOKE measurements (\SI{\pm 643}{\milli\tesla}) or out-of-plane for the P-MOKE measurements (\SI{\pm 102}{\milli\tesla}), respectively. 
Fig.~\ref{fig:setup}a shows a schematic of the VIS-MOKE configuration, illustrating the longitudinal (orange) and collinear polar (blue) geometries together with the corresponding orientations of the applied magnetic fields. 
The Kerr rotation ($\theta_\text{K}$) was measured by splitting the reflected probe beam into two orthogonally polarized components using a Wollaston prism, which are detected by a boxcar-gated balanced photodiode pair. 
The magnetic contrast was achieved by taking the difference of two measurements for opposite magnetic field directions ($S^{\uparrow,\downarrow}(t)$) and normalizing this difference to the unpumped signal $S^{\uparrow,\downarrow}(t<t_0)$:
\[
\frac{\theta_\text{K}(t)}{\theta_\text{K}(t<t_0)}=
\frac{S^{\uparrow}(t)-S^{\downarrow}(t)}
     {S^{\uparrow}(t<t_0)-S^{\downarrow}(t<t_0)}
=
\frac{M(t)}{M(t<t_0)} .
\]

The element-specific, time-resolved XUV-MOKE measurements were performed using \SI{\leq 25}{\femto\second} short XUV pulses with photon energies from 45 to \SI{72}{\electronvolt}, produced by a high-harmonic generation (HHG) source. \cite{yao_tabletop_2020}
The analysis was carried out by integrating over the HHG peak at \SI{60.4}{\electronvolt}, resonant with the Co M$_{3,2}$ resonance \cite{willems_magneto-optical_2019}. 
The HHG process was driven by focusing near-infrared pulses (\SI{800}{\nano\meter} central wavelength, \SI{25}{\femto\second} pulse duration, \SI{3}{\kilo\hertz} repetition rate) into a neon-filled gas cell. 
A fraction of this near-infrared beam is coupled out and used for optical excitation of the sample, leading to a temporal resolution of \SI{\approx 35}{\femto\second}.
The measurements were carried out in reflection geometry using transverse (XUV-T-MOKE) and polar (XUV-P-MOKE) configurations (see Fig.\ref{fig:setup}b). 
In the XUV-T-MOKE geometry, the magnetization was oriented perpendicular to the plane of incidence of p-polarized light, providing sensitivity only to the in-plane magnetization component~\cite{la-o-vorakiat_ultrafast_2009}. 
In the XUV-P-MOKE geometry, the magnetic field was applied along the surface normal, yielding dominant sensitivity to the out-of-plane magnetization component \cite{richter_spectroscopic_2025}.
The XUV polarization angle $\phi$ was controlled by rotating a $\lambda/2$ wave plate in the near-infrared beam upstream of the HHG cell. 
After reflection by the sample, the XUV radiation was spectrally dispersed by a flat-field grating and detected on an in-vacuum CCD camera. 
The angle of incidence was fixed at $\theta = 45^\circ$, and, similar to the VIS-MOKE scheme, two measurements are taken for an alternating magnetic field of \SI{\pm 75}{\milli\tesla}.
The magnetic contrast is then evaluated by the asymmetry between the two measurements, defined as
\begin{equation}
A_{\phi}(t) = \frac{R^{\uparrow}(t) - R^{\downarrow}(t)}{R^{\uparrow}(t) + R^{\downarrow}(t)}\propto M(t),
\end{equation}
where $R^{\uparrow,\downarrow}(t)$ denote the reflected intensities for opposite magnetization directions. 

\section{VIS-MOKE measurements}\label{sec:VIS}


\begin{figure}
    \includegraphics[width=\linewidth]{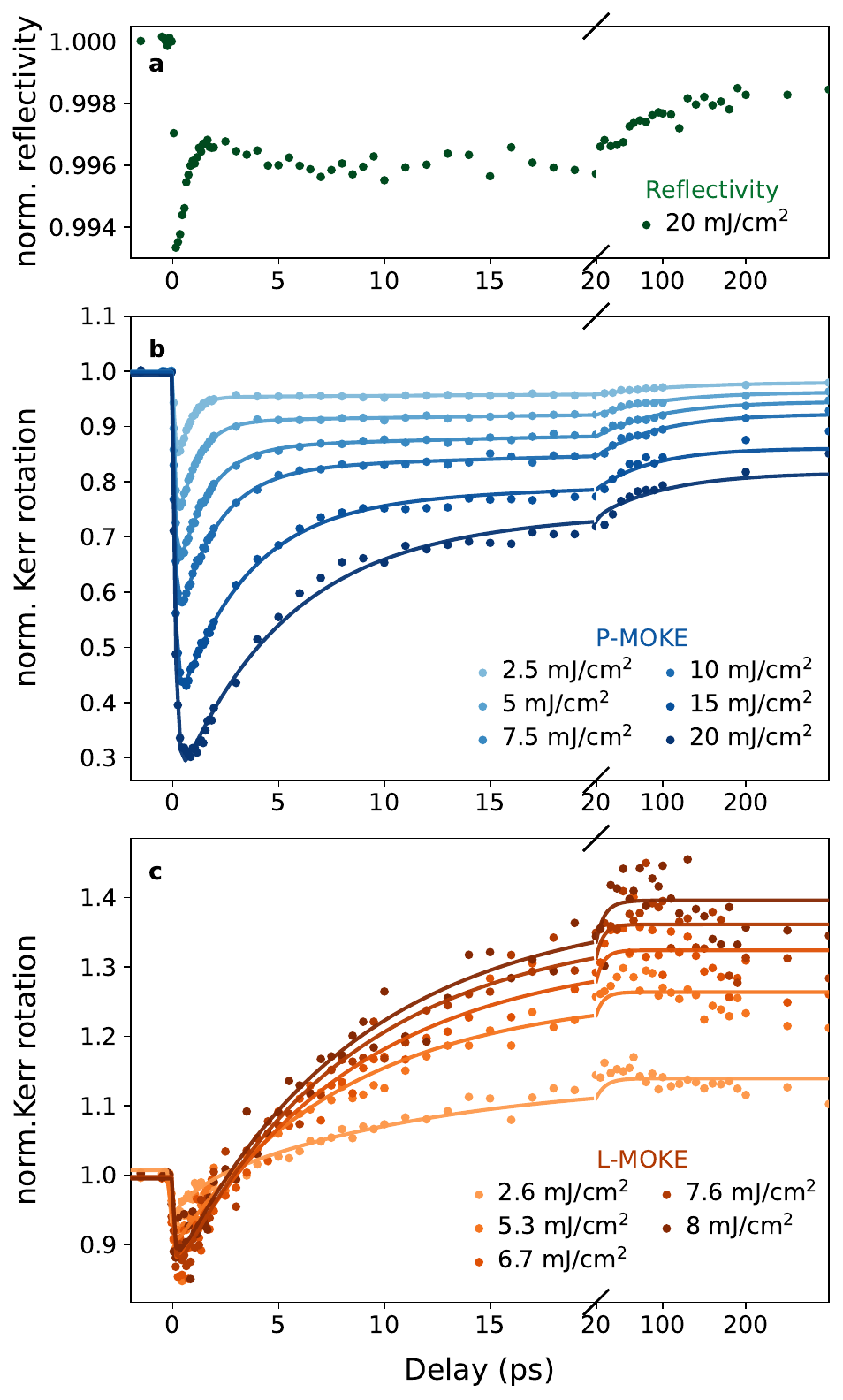}
    \caption{Time-resolved non-magnetic reflectivity (\textbf{a}) and VIS-MOKE measurements (\textbf{b}, \textbf{c}) performed with an \SI{800}{\nano\meter} pump and a \SI{400}{\nano\meter} probe wavelength in P-MOKE (b) and L-MOKE (c) geometry.
    The solid lines denote the respective convoluted triple exponential fits.}
    \label{fig:opticalMOKE2}
\end{figure}

Fig.~\ref{fig:opticalMOKE2} shows representative time traces recorded in the visible-light P-MOKE and L-MOKE geometries, as well as the transient, non-magnetic reflectivity change for reference. 
The VIS-MOKE time traces (panel b and c) were analyzed using a phenomenological fit model consisting of a sum of three exponential trends, convolved with a Gaussian instrument response function:
\begin{equation}
\begin{split}
A(t)&=y_0+\sum_{i=1}^{3} C_i\Big[\Theta(t-t_0)\big(1-e^{-(t-t_0)/\tau_i}\big)\Big]\ast G_\sigma(t),\\
G_\sigma(t)&=\frac{1}{\sqrt{2\pi}\sigma}\exp\left(-\frac{t^2}{2\sigma^2}\right).
\end{split}
\end{equation} \label{eq:triple_fit}
Here, $y_0$ denotes the MOKE signal before excitation ($t<t_0$), the Heaviside function $\Theta(t-t_0)$ enforces causality, and the convolution with $G_\sigma(t)$ accounts for the finite temporal resolution of the experiment ($\SI{\approx 50}{\fs}$). 
The parameters $C_i$ and $\tau_i$ describe the amplitudes and time constants of the three exponential trends, respectively, with the index $i$ labeling the different dynamical contributions: $i=1$ denotes the demagnetization process, $i=2$ the fast remagnetization, and $i=3$ the slower recovery and spin reorientation processes observed in the P-MOKE and L-MOKE data, respectively.
The demagnetization and remagnetization time constants extracted from the P-MOKE measurements were used to constrain the parameter space for the corresponding L-MOKE fits.

In the P-MOKE geometry (see Fig.~\ref{fig:opticalMOKE2}b), the normalized Kerr signal exhibits an ultrafast reduction immediately after optical excitation at $t_0$, corresponding to the laser-induced demagnetization of the DyCo$_5$ layer, with the demagnetization amplitude increasing as a function of incident excitation fluence. 
This initial ultrafast demagnetization with a time constant of $\tau_{\mathrm{1}}=(0.116 \pm 0.008)\,\mathrm{ps}$ is followed by a two-step recovery process, consisting of a fast remagnetization with $\tau_{\mathrm{2}}=(1.6 \pm 1.0)\,\mathrm{ps}$ and a subsequent slower recovery with $\tau_{\mathrm{3}}=(82 \pm 13)\,\mathrm{ps}$.
However, the magnetic signal does not recover to its pre-excitation level within the measured time range of \SI{300}{ps} after excitation, pointing to a long-lived thermally induced demagnetization component of up to $\approx 20$\% for the highest excitation fluence shown. 
Note that these time constants are averaged over all investigated fluences, since they are independent of the fluence within the experimental uncertainty.
The corresponding L-MOKE data reveal a similar ultrafast demagnetization and remagnetization step on the early timescales, with averaged time constants of $\tau_{\mathrm{1}}=(0.14 \pm 0.07)\,\mathrm{ps}$ and $\tau_{\mathrm{2}}=(6.6 \pm 2.8)\,\mathrm{ps}$.

However, in contrast to the P-MOKE response, the L-MOKE signal subsequently overshoots the equilibrium value by approximately $10$--$40\%$ with a time constant of $\tau_{\mathrm{3}}=(15 \pm 5)\,\mathrm{ps}$.
This pronounced overshoot indicates a substantial transient enhancement of the in-plane magnetization component and is consistent with a laser-induced modification of the magnetic anisotropy toward an in-plane easy axis.
The data further reveals that the amplitude of the overshoot depends on the incident fluence, indicating a stronger transient tilt of the magnetization toward the in-plane direction for stronger excitation.  
The reflectivity signal shown in Fig.~\ref{fig:opticalMOKE2}a confirms that the observed dynamic rise of the L-MOKE signal cannot be explained by a simple change in the sample’s reflectivity, since, following the initial ultrafast heating and thermalization phase, only a negligible reflectivity change occurs on the relevant timescales of a few tens to a few hundreds of picoseconds.
However, a quantitative interpretation and discussion based solely on the VIS-MOKE data is difficult, as the L-MOKE probing angle still allows for a significant P-MOKE contribution to the Kerr signal, mixing the dynamics of the in-plane and out-of-plane magnetization components. 
This effect is particularly pronounced during the SRT, where the magnetization evolves between out-of-plane and in-plane orientations. 
In this regime, the longitudinal and polar Kerr signals become mixed with a field-dependent weight, which complicates a quantitative analysis of the data~\cite{ding_experimental_2000}.
Furthermore, even a slight inhomogeneity and misalignment of the applied alternating magnetic field with respect to the sample plane causes a field component along the sample normal, effectively generating the geometry of a P-MOKE measurement.
This behavior is observed for incident fluences above \SI{10}{\flu}, where the laser-induced heating of the sample quenches the coercive field, causing a strong P-MOKE contrast due to field-induced toggling of the out-of-plane magnetization that overlays the L-MOKE measurements.

To directly access the in-plane component without the cross-sensitivity, we employ element-specific XUV-MOKE measurements, where the transverse geometry provides direct and exclusive sensitivity to the in-plane magnetization component~\cite{richter_spectroscopic_2025,Oppeneer_2001}.

\section{XUV-MOKE simulations}

To enable a quantitative analysis of the XUV-MOKE experiments, we first perform comparative simulations at the Co M$_{3,2}$ resonance at \SI{60.4}{\electronvolt} using the wave-propagation matrix code \textsc{udkm1Dsim} \cite{schick_udkm1dsim_2021,elzo_x-ray_2012}. 
This allows for a direct and quantitative comparison between the simulated magneto-optical responses and the experimental data.
As an input for the simulations, we use the structural parameters obtained via XRR (see Fig.~\ref{fig:xraydiff} and Table~\ref{tab:simulation_parameters}), the resonant magneto-optical dielectric functions from \textcite{willems_magneto-optical_2019} and the off-resonant form factors from the \textcite{henke_x-ray_1993} database.

Motivated by the theoretical description of thermally induced canting caused by a temperature-dependent magnetic anisotropy~\cite{donges_magnetization_2017}, we consider the magneto-optical response of a magnetization vector tilted away from the out-of-plane direction. 
Such a tilt naturally gives rise to simultaneous out-of-plane and in-plane magnetization components, which contribute differently to polar and transverse MOKE.
While the XUV-P-MOKE signal is primarily sensitive to the out-of-plane component of the magnetization, the XUV-T-MOKE response scales exclusively with the in-plane component perpendicular to the plane of incidence. 
A progressive canting of the magnetization toward the film plane therefore manifests as a reduction of the XUV-P-MOKE signal and a corresponding increase of the XUV-T-MOKE amplitude. 
The magnetic moment can also decrease as a function of temperature, reducing the overall magneto-optical response. 
This almost complementary behavior provides a direct handle on the magnetization orientation during the SRT.

\begin{figure}
    \centering
    \includegraphics[width=\linewidth]{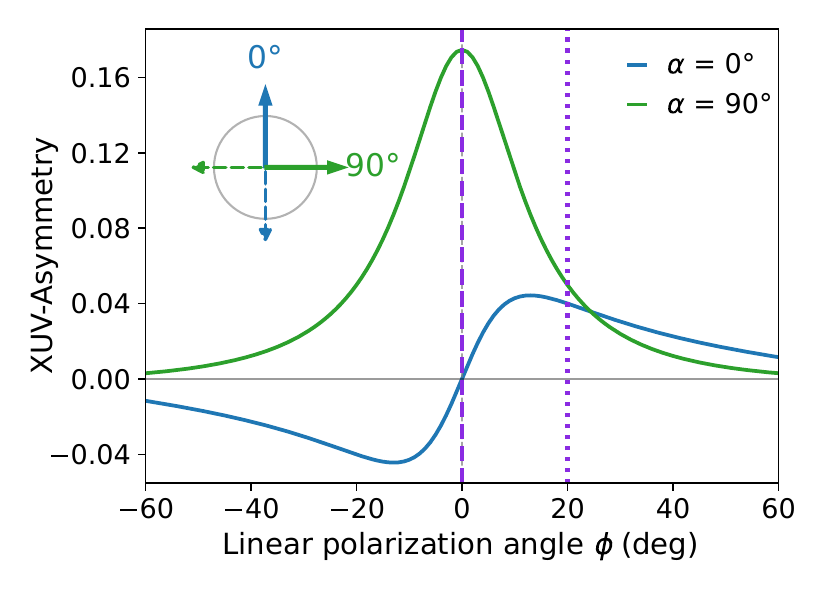}
    \caption{Simulation of the XUV-P-MOKE asymmetry for out-of-plane ($\alpha$ = \SI{0}{\degree}) and the XUV-T-MOKE asymmetry for in-plane ($\alpha$ = \SI{90}{\degree}) magnetization as a function of the polarization angle $\phi$ of the incident XUV probing beam.}
    \label{fig:simpol}
\end{figure}

We will first look at the XUV-P-MOKE and XUV-T-MOKE responses in the limiting cases of pure out-of-plane or in-plane magnetization, respectively.
Fig.~\ref{fig:simpol} shows magnetic asymmetries as a function of the XUV polarization angle $\phi$ for two perpendicular magnetization orientations described by the canting angle $\alpha$, where $\alpha=\SI{0}{\degree}$ corresponds to pure out-of-plane magnetization measured by XUV-P-MOKE and $\alpha=\SI{90}{\degree}$ to pure in-plane magnetization measured by XUV-T-MOKE. 
The photon energy of the incident XUV radiation is set to \SI{60.4}{\electronvolt}, and $\phi=\SI{0}{\degree}$ corresponds to p-polarization.
The asymmetry is calculated from the simulated reflectivities for opposite magnetization directions along the out-of-plane and in-plane directions (solid and dashed arrows in the inset of Fig.~\ref{fig:simpol}), respectively, corresponding to the experimental alignment of the toggled magnetic field.
In the first case, the resulting polarization dependent asymmetry shows the characteristic XUV-P-MOKE response, exhibiting a zero crossing for p-polarization ($\phi=\SI{0}{\degree}$). 
In the second case, it corresponds to a typical XUV-T-MOKE signal with a Lorentzian-shaped maximum centered around $\phi=\SI{0}{\degree}$.
The calculations predict that in transverse geometry, a complete in-plane alignment of the magnetization would lead to an asymmetry of $\approx 17.5\%$ at the Co M$_{3,2}$ resonance when the XUV pulses are fully p-polarized ($\phi=\SI{0}{\degree}$, indicated by the dashed purple line).
In polar geometry, respectively, the asymmetry is predicted to peak at $\approx 4$\% when the magnetization is fully aligned along the out-of-plane direction and the XUV polarization is slightly rotated away from the p-polarization axis. 
In the experiment, the incident polarization was set to $\phi\approx\SI{20}{\degree}$ (dotted purple line), which has been identified as the best compromise between asymmetry and increased reflectivity for a polarization tuned away from Brewster's angle.

\begin{figure}
    \centering
    \includegraphics[width=\linewidth]{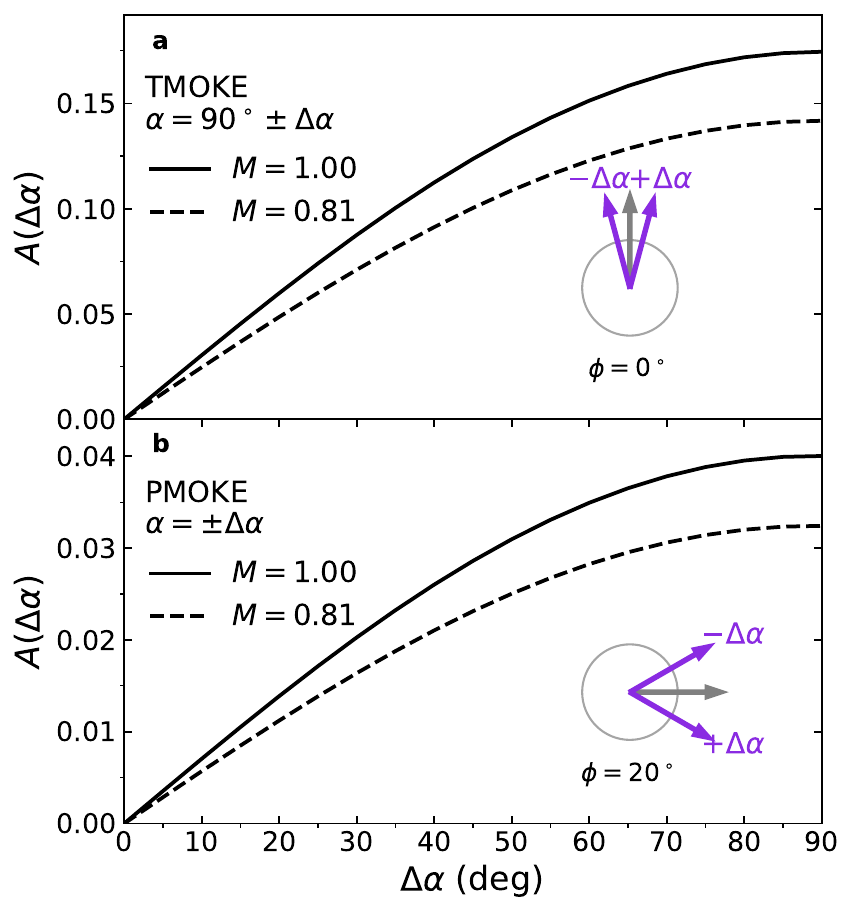}
    \caption{Simulations of the asymmetry $A$ as a function of the canting angle $\Delta\alpha$ for \textbf{a} an increasing in-plane component measured by XUV-T-MOKE and \textbf{b} an increasing out-of-plane component measured by XUV-P-MOKE. 
    }
    \label{fig:SIM} 
\end{figure}

We now quantify the asymmetry change when the magnetization is tilted with respect to pure out-of-plane and in-plane directions.
Depicted in Fig.~\ref{fig:SIM}, we show the dependence of the XUV-T-MOKE and XUV-P-MOKE asymmetries on the magnetization orientation, by evaluating the asymmetry as a function of the canting angle $\Delta\alpha$. 
The asymmetry is defined as
\begin{equation}
A(\Delta\alpha)=\frac{R_\uparrow(\Delta\alpha)-R_\downarrow(-\Delta\alpha)}{R_\uparrow(\Delta\alpha)+R_\uparrow(-\Delta\alpha)},
\end{equation}
where $R_\uparrow(\Delta\alpha)$ and $R_\downarrow(-\Delta\alpha)$ denote the reflected intensities for opposite magnetization tilt with respect to the initial geometry. 

Fig.~\ref{fig:SIM}a shows the simulated asymmetry $A(\Delta\alpha)$ for the out-of-plane to in-plane tilt obtained in the XUV-T-MOKE configuration (evaluated at $\phi$=0). 
Note that this also corresponds to the SRT process where the magnetization is initially oriented along the out-of-plane direction and then rotates by $\pm \Delta\alpha$ toward the sample plane, with a direction defined by the toggling in-plane magnetic field. 
As expected, the predicted XUV-T-MOKE asymmetry increases with the projection of the magnetization vector along the probed in-plane direction. 

Fig.~\ref{fig:SIM}b shows the opposite case of a rising out-of-plane component in XUV-P-MOKE configuration, yielding the same angle dependence with overall smaller asymmetry values in line with the simulations depicted in Fig.~\ref{fig:simpol}. 
The complementary angular dependence of $A(\Delta\alpha)$ in XUV-P-MOKE and XUV-T-MOKE provides a direct mapping between the measured asymmetry amplitudes and the magnetization orientation. 
This allows us to interpret the experimentally observed pump-induced changes in the asymmetry in terms of a continuous canting of the magnetization from the out-of-plane toward the in-plane direction.

\section{Polarization-resolved measurements in XUV-P-MOKE and XUV-T-MOKE geometry}

\begin{figure} 
    \centering
    \includegraphics[width=\linewidth]{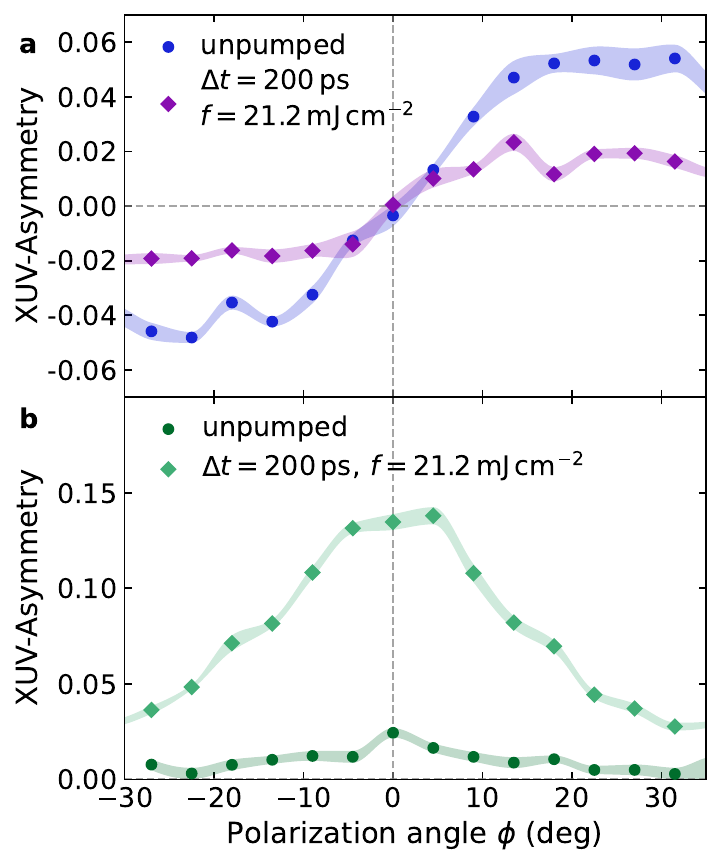}
    \caption{Polarization-dependent measurements in \textbf{a} XUV-P-MOKE and \textbf{b} XUV-T-MOKE geometry, comparing the unexcited (unpumped, circles) to the excited state (squares) recorded at a pump-probe delay of $\Delta t$=\SI{200}{\ps} for an incident excitation fluence of \SI{21.2}{\flu}.
    In both polar and transverse geometries, a toggling magnetic field of \SI{\pm 75}{\milli\tesla} was applied either along the out-of-plane or in-plane direction, respectively.
    For the XUV-P-MOKE measurement in the excited state (purple squares in panel a), a constant in-plane field of \SI{75}{mT} was applied in addition to the toggling out-of-plane field.
    The shaded areas correspond to the standard error of the mean ($68.2\%$ confidence interval).}
    \label{fig:pol}
\end{figure}

To confirm and quantify the SRT dynamics, we perform polarization-dependent XUV-MOKE measurements. 
Fig.~\ref{fig:pol}a shows the XUV-P-MOKE asymmetry as a function of polarization angle, comparing the unexcited (unpumped) state to the reoriented state after excitation with an incident fluence of \SI{21.2}{\milli\joule\per\centi\meter\squared} recorded at a pump-probe delay of \SI{200}{\pico\second}.
The data exhibits the characteristic zero crossing of the P-MOKE response at p-polarization. 
In the unpumped state, the extrema reach values of $\pm5.4\,\%$, which, within the experimental uncertainty, is close to the values predicted by the simulations shown in Fig.~\ref{fig:simpol}. 
Following optical excitation, the asymmetry is reduced.
Note, that for this measurement, the out-of-plane magnetic field toggled for the P-MOKE measurement was supplemented by an additional in-plane field of \SI{75}{mT}, which has negligible effect on the asymmetry in the unexcited state, but strongly reduces it in the excited state compared to a pure out-of-plane field (not shown).
This implies that the reduction of the asymmetry is not only caused by demagnetization, but also by a field-induced tilting of the magnetization direction toward the plane, enabled by a significantly reduced anisotropy after excitation.

Fig.~\ref{fig:pol}b shows the same measurements obtained in the XUV-T-MOKE configuration. 
In the unpumped state, the asymmetry at p-polarization ($\phi=$\SI{0}{\degree}) is almost zero, consistent with the simulations of predominantly out-of-plane magnetization. 
Note that a non-vanishing asymmetry is reasonable even in the unpumped state, as the applied in-plane magnetic field slightly pulls the magnetization vector towards the sample plane. 
When excited with an incident fluence of \SI{21.2}{\milli\joule\per\centi\meter\squared}, the asymmetry increases to $0.14\pm0.01$ at a pump-probe delay of \SI{200}{\pico\second}.
This pronounced increase directly evidences the development of a substantial in-plane magnetization component, fully consistent with the canting scenario inferred from the polarization-resolved XUV-P-MOKE measurements and the simulations.

\section{Ultrafast magnetization reorientation from out-of-plane to in-plane}

\begin{figure} 
    \includegraphics[width=\linewidth]{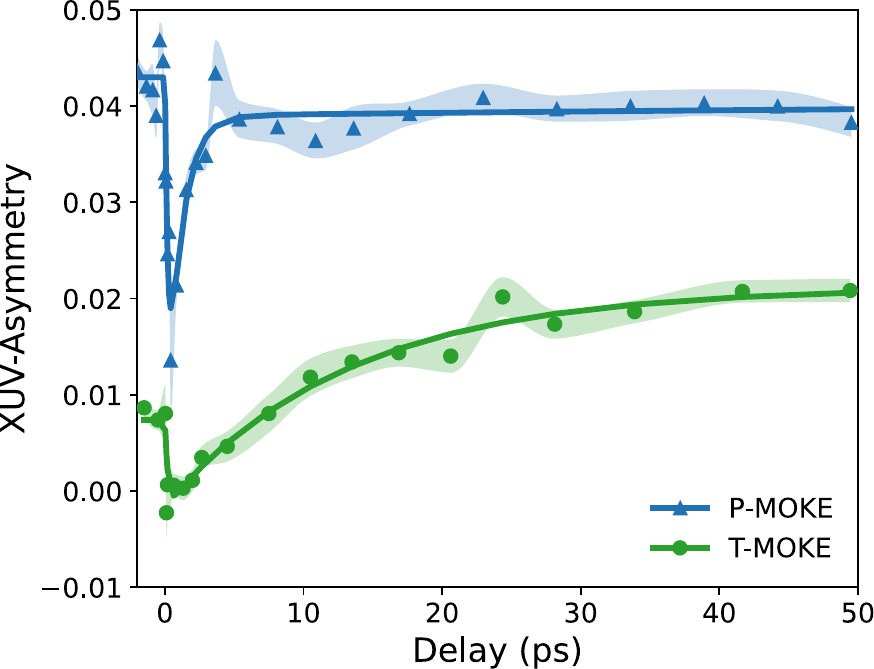}
    \caption{Time-resolved XUV-P-MOKE (blue triangles) and XUV-T-MOKE (green circles) measurements at \SI{60.4}{\electronvolt} photon energy upon excitation with \SI{14.1}{\flu} incident fluence.
    The lines correspond to triple exponential fits, the shaded areas to the standard error of the mean ($68.2\%$ confidence interval).}
    \label{fig:ultra}
\end{figure}

In this section, we discuss the ultrafast response of the magnetic asymmetry measured in XUV-T-MOKE and XUV-P-MOKE geometries. 
Fig.~\ref{fig:ultra} shows the time-resolved asymmetry for a photon energy of \SI{60.4}{\electronvolt} upon excitation with an incident excitation fluence of \SI{14.1}{\flu}. 
The time-resolved asymmetry data is fitted using the Gaussian-convolved multi-exponential model shown in Eq.~\ref{eq:triple_fit}.

For consistency, the time constants $\tau_1$ and $\tau_2$ are kept identical for the P-MOKE and T-MOKE fits, as the intrinsic speed of the ultrafast demagnetization and fast remagnetization dynamics are not expected to change with respect to the measured component of the magnetization vector.

Before excitation (delay \SI{<0}{ps}), the XUV-P-MOKE asymmetry is $(4.3 \pm 0.1)\,\%$, consistent with the simulated and measured P-MOKE asymmetry shown in Figs.~\ref{fig:simpol} and \ref{fig:pol}a. 
Following optical excitation, the sample undergoes an ultrafast demagnetization and subsequent remagnetization along the out-of-plane direction.
The corresponding time constants are $\tau_1 = (0.212 \pm 0.098)\,\text{ps}$, $\tau_2 = (1.0 \pm 0.6)\,\text{ps}$, which are, within the experimental uncertainty, similar to the respective values obtained from the polar VIS-MOKE measurements. 
In the following, the asymmetry recovers to $(4.0 \pm 0.4)\%$ at a delay of \SI{50}{\pico\second}, corresponding to $\approx 93\%$ of its initial value.
This remaining offset indicates a long-lived reduction of the out-of-plane magnetization due to an elevated sample temperature. 
Similar temperature-induced reductions of the Co asymmetry have previously been observed in static XMCD measurements \cite{donges_magnetization_2017}.

In the XUV-T-MOKE geometry, the asymmetry before excitation is $(0.74 \pm 0.14)$\%. 
Systematic measurements performed for different alternating magnetic field amplitudes (not shown) reveal a linear scaling of this unpumped asymmetry with the field amplitude, which is applied along the initial hard axis and drags the magnetization slightly toward the plane.
At $t_0$ and similar to the measurements obtained in XUV-P-MOKE geometry, the data shows an ultrafast demagnetization and subsequent fast remagnetization along the in-plane direction.
In contrast to the XUV-P-MOKE response, the XUV-T-MOKE asymmetry overshoots the value of the initial unexcited state, reaching $(2.08 \pm 0.12)$\% at \SI{50}{\pico\second} delay.  
The large enhancement of the T-MOKE asymmetry demonstrates the formation of an in-plane magnetization component, which is not yet reaching equilibrium within the scanned range of \SI{50}{\pico\second}. 
Comparing the XUV-P-MOKE and XUV-T-MOKE response further reveals that the observed spin reorientation starts to dominate the dynamics on timescales where the magnetization has mostly recovered (\SI{>2}{ps}).

\begin{figure}
    \centering
    \includegraphics[width=\linewidth]{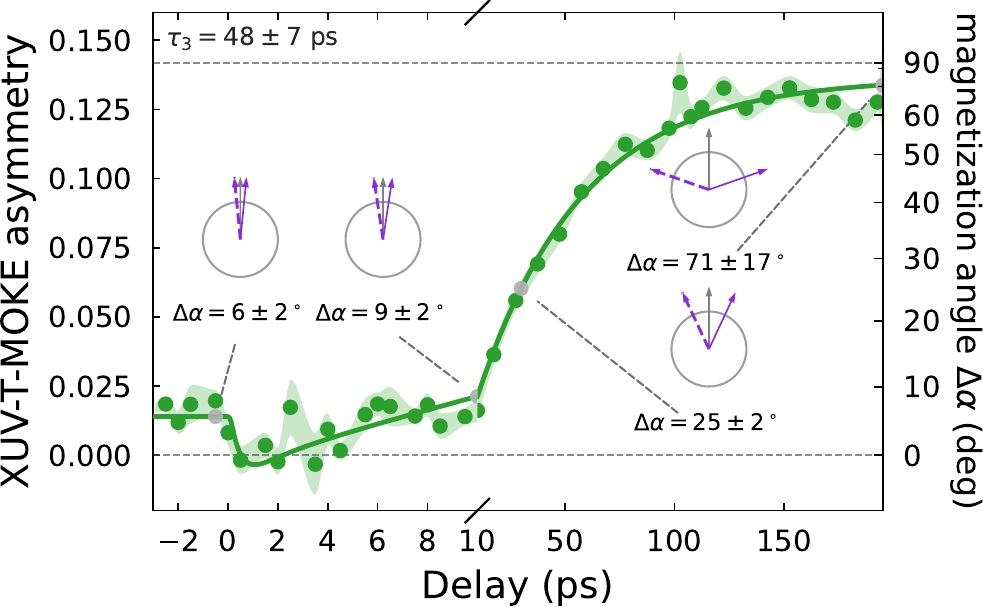}
    \caption{Time-resolved XUV-T-MOKE measurements (green circles) at \SI{60.4}{\electronvolt} photon energy upon excitation with an incident fluence of \SI{21.2}{\flu}, applying an alternating magnetic field of \SI{\pm75}{\milli \tesla}. 
    The solid line corresponds to a triple exponential fit and the shaded areas denote the standard error of the mean ($68.2\%$ confidence interval).
    By comparing the measured asymmetry to the simulated values, the T-MOKE data can be related to the spin canting angle $\alpha$.}
    \label{fig:delay200_AsymPhi}
\end{figure}

For an accurate determination of the SRT time constant $\tau_3$, additional time-resolved measurements are performed, covering a larger delay range (see Fig.~\ref{fig:delay200_AsymPhi}).
The excitation fluence for this long-range scan was chosen according to a fluence dependent characterization of the asymmetry enhancement at late timescales, showing saturation for an incident fluence of \SI{21.2}{\flu}. 
The data reveals that the in-plane canting progresses over long timescales of up to \SI{200}{\pico\second} with a characteristic SRT time constant of $\tau_{3} = (48 \pm 7) \si{\pico\second}$.
Note, that the obtained value of $\tau_{3}$ is significantly larger than the corresponding time constant obtained from the longitudinal VIS-MOKE measurements ($\tau_{\mathrm{3}}=(15 \pm 5)\,\mathrm{ps}$).

To evaluate the time dependence of the magnetization vector tilt, we map the time-resolved XUV-T-MOKE asymmetry to the magnetization canting angle ($\Delta \alpha$) according to the relation shown in Fig.~\ref{fig:SIM}a.
As the sample heating induced by the excitation results in a reduction of both out-of-plane and in-plane components and the respective asymmetries, we scale the simulated asymmetry values by $81\%$, corresponding to the persistent demagnetization at \SI{200}{ps} observed in the polar VIS-MOKE measurements for a comparable excitation fluence of \SI{20}{\flu} (compare Fig.~\ref{fig:opticalMOKE2} and the dashed line in Fig.~\ref{fig:SIM}). 
This calibration of the simulations accounts for the thermally elevated state of the sample, relating a maximum XUV-T-MOKE asymmetry of $\approx 14.1$\% to a complete in-plane tilt of \SI{90}{\degree}.
The analysis reveals that the finite asymmetry before excitation corresponds to a spin tilt of $(6 \pm 2)$\si{\degree}, caused by the applied magnetic field, which increases progressively during the observed SRT to a tilt angle of approximately $(71 \pm 17)$\si{\degree}.

\section{Discussion}

The observation of a substantial increase of the in-plane magnetization component evolving on timescales of tens of picoseconds as evidenced by both the VIS-L-MOKE and XUV-T-MOKE measurements clearly shows that an ultrafast optical excitation is able to drive a transient spin reorientation in DyCo$_5$.
The observed reorientation evolves with characteristic time constants in the range of $\SIrange{\approx 15}{\approx 50}{ps}$, which fall into the same order of magnitude as typical spin-lattice relaxation times in RE elements \cite{Wietstruk2011,Eschenlohr2014}, pointing at the transient heating of the lattice as responsible driver for the SRT process.
A plausibility check based on simple heat diffusion simulations, assuming a single heat bath to estimate the longevity of the transient, laser-induced heating along the depth of the studied sample, predicts that an excitation of \SI{21.2}{\flu} indeed leads to elevated lattice temperatures exceeding the reported SRT range of \SIrange{320}{360}{\kelvin}~\cite{donges_magnetization_2017} for sufficiently long timescales of \SI{200}{\pico\second}.

To better understand the dynamics of the SRT, we have to recall that the laser-driven evolution of the magnetization in DyCo$_5$ can be understood as a competition between the applied magnetic field and transient changes in the sublattice anisotropies. 
In equilibrium, the strong Dy anisotropy confines the spin axis to the out-of-plane direction. 
The laser excitation heats the magnetic layer, rapidly reducing the Dy anisotropy until the Co anisotropy eventually starts to dominate, driving the system toward an in-plane easy axis.
We expect the anisotropy changes to be ultrafast, such that following the rapid remagnetization ($\tau_2 < \SI{2}{ps}$), the spin system begins to realign along the new in-plane easy axis, typically described by the Landau–Lifshitz–Gilbert equation.
As such, both the excitation fluence -- setting the effective strength and direction of the transient anisotropy -- and the external magnetic field is expected to govern the timescale of the SRT.
This is in line with the distinct SRT time constants observed in the VIS- and XUV-MOKE data, which can be attributed to the different experimental conditions and thermal regimes transiently reached during the visible and XUV light measurements.
In particular, the highest fluence usable in the presented VIS-MOKE measurements in L-MOKE geometry is limited to \SI{10}{\flu}, as for higher fluences the strong cross-sensitivity to the out-of-plane magnetization due to an increasing P-MOKE contribution prevents a clear measurement of the in-plane component.
The significantly higher saturation fluence of the SRT observed in the XUV measurements suggests that the lower fluence used in the VIS-MOKE measurements is not sufficient to fully drive the DyCo$_5$ layer into the SRT regime on timescales long enough for the spins to fully reorient along the new in-plane anisotropy direction.
Consequently, the system cools down below the SRT regime while the reorientation has not yet been completed, quenching the SRT dynamics with the onset of relaxation, which effectively shortens the retrieved time constant.
In contrast, the XUV measurements were carried out using excitation of nearly twice the fluence and repetition rate (\SI{3}{\kilo\hertz} vs.~\SI{1.5}{\kilo\hertz}).
In addition to a reduced cooling rate caused by the vacuum environment needed for experiments in the XUV spectral range, the sample is expected to reach significantly higher equilibrium and transient temperatures, facilitating the long-lived excitation of the full probed volume of the DyCo$_5$ layer into the SRT regime.
Note that the proposed field dependence could also contribute to the significantly faster SRT dynamics (\SI{\approx 15}{ps}) observed in the VIS-L-MOKE measurements (see Fig.~\ref{fig:opticalMOKE2}c), which were recorded using a much larger applied field of \SI{643}{mT} compared to \SI{75}{mT} in the XUV-T-MOKE measurements, potentially accelerating the spin reorientation process.

Beyond the experimental conditions, the different sublattice sensitivities of the visible and XUV probes may also affect the observed time constants.
While the polarity of magnetic hysteresis loops measured using visible light (not shown) suggests a predominantly Co-sensitive response, contributions from Dy $5d$ electrons and unoccupied Dy $4f$ states to the magneto-optical transitions cannot be excluded~\cite{lang_study_1981}. 
In contrast, the XUV-MOKE measurements at the Co $M_{3,2}$ resonance are exclusively sensitive to the Co sublattice magnetization, making the two probes inherently non-equivalent.
This can be significant in case of sublattice-specific SRT dynamics, potentially caused by the slightly non-collinear alignment between the magnetic moments of Co and Dy~\cite{donges_magnetization_2017}, as well as their different demagnetization and remagnetization timescales, which can lead to significant differences in the magnetization up to \SI{50}{\pico\second} after excitation~\cite{abrudan_element-specific_2021}.

In order to disentangle the sublattice contributions and identify the microscopic processes defining the timescales of the SRT, further systematic measurements of the SRT dynamics as a function of both excitation fluence and applied field are required to map the transient anisotropy as it evolves following laser excitation. 
Future work will pursue this by combining element-resolved XUV spectroscopy 
with atomistic spin dynamics simulations, to provide a complete characterization of the anisotropy landscape during the laser-driven SRT.

\section{Conclusion}

We have investigated the ultrafast spin-reorientation dynamics in ferrimagnetic DyCo$_5$ thin films by combining optical MOKE with element-specific XUV-MOKE measurements carried out at the Co M$_{3,2}$ resonance.
The time-resolved data provide clear evidence of a spin reorientation on picosecond timescales, demonstrating that an optical excitation is sufficient to transiently drive the system into the temperature regime associated with the SRT.
By analyzing direction-selective measurements in polar and transverse geometry, the XUV-MOKE measurements allow a quantitative analysis of the in-plane tilting of the Co spins, as well as the extraction of the characteristic SRT time constant.
While the present measurements provide evidence for a transient modification of the magnetic anisotropy landscape leading to a spin reorientation of the Co sublattice, the exact role of the heavy rare-earth Dy constituent remains an open question. 
Future element-specific measurements of the Dy sublattice dynamics, employing the combined polar and transverse XUV-MOKE approach at the Dy~N$_{5,4}$ resonances~\cite{hennecke_ultrafast_2022, hennecke_transient_2025}, will allow us to fully disentangle the microscopic pathways underlying the ultrafast SRT in rare-earth–transition-metal alloys.

\section*{Data availability statement}
The data and further details on the analysis are available from the authors upon reasonable request. 

\begin{acknowledgments}
C.v.K.S., J.R. and S.E. acknowledge financial support from the German Research Foundation (DFG, Germany) through CRC/TRR 227 project A02 (project ID 328545488). 
M.S. acknowledges funding from the European Union via ERDF project 1.8/15.
The samples have been grown by magnetron sputtering at the Helmholtz-Zentrum Berlin.
\end{acknowledgments}
\bibliography{ref}
\end{document}